\documentclass{article}
\usepackage{ijcai23}

\usepackage{xcolor}
\usepackage{times}
\usepackage{soul}
\usepackage{url}
\usepackage[hidelinks]{hyperref}
\usepackage[utf8]{inputenc}
\usepackage[small]{caption}
\usepackage{graphicx}
\usepackage{amsmath}
\usepackage{amsthm}
\usepackage{booktabs}
\usepackage{algorithm}
\usepackage{algorithmic}
\usepackage[switch]{lineno}
\usepackage[utf8]{inputenc}
\usepackage{longtable}
\usepackage{enumitem}
\usepackage{tcolorbox}
\usepackage[utf8]{inputenc}
\usepackage{CJKutf8} 
\usepackage{tcolorbox}
\usepackage{xcolor}
\definecolor{newsColor}{rgb}{0.0, 0.5, 0.0}
\definecolor{financeColor}{rgb}{0.0, 0.0, 1.0}
\definecolor{stockPriceColor}{rgb}{1.0, 0.0, 0.0}
\tcbuselibrary{skins}
\usepackage{mathtools}
\usepackage{amssymb}

\usepackage[utf8]{inputenc}
\usepackage{fancyhdr}

\nolinenumbers

\title{FinRobot: An Open-Source AI Agent Platform \\ for Financial Applications using Large Language Models}

\author{
    Hongyang (Bruce) Yang$^{1}$, Boyu Zhang$^{1}$, Neng Wang$^{1}$, Cheng Guo$^{1}$, \\Xiaoli Zhang$^{1}$, Likun Lin$^{1,2}$, Junlin (Jason) Wang $^{1,4}$,Tianyu Zhou$^{1}$, Mao Guan$^{1}$,\\ Runjia (Luna) Zhang$^{1,4}$, Christina Dan Wang$^{1,3}$\thanks{Corresponding author};
    \affiliations
    $^1$AI4Finance Foundation; $^2$Columbia University;\\$^3$Shanghai Frontiers Science Center of Artificial Intelligence and Deep Learning, \\NYU Shanghai; Business Division, NYU Shanghai; \\$^4$Shanghai AI-Finance School ECNU; 
    \emails
    contact@ai4finance.org; 
}

\begin{document}

\maketitle

\begin{abstract}
As financial institutions and professionals increasingly incorporate Large Language Models (LLMs) into their workflows, substantial barriers, including proprietary data and specialized knowledge, persist between the finance sector and the AI community. These challenges impede the AI community's ability to enhance financial tasks effectively. Acknowledging financial analysis's critical role, we aim to devise financial-specialized LLM-based toolchains and democratize access to them through open-source initiatives, 
promoting wider AI adoption in financial decision-making.

In this paper, we introduce FinRobot, a novel open-source AI agent platform supporting multiple financially specialized AI agents, each powered by LLM. Specifically, the platform consists of four major layers: 
1) the Financial AI Agents layer that formulates Financial Chain-of-Thought (CoT) by breaking sophisticated financial problems down into logical sequences; 
2) the Financial LLM Algorithms layer dynamically configures appropriate model application strategies for specific tasks;
3) the LLMOps and DataOps layer produces accurate models by applying training/fine-tuning techniques and using task-relevant data;
4) the Multi-source LLM Foundation Models layer that integrates various LLMs and enables the above layers to access them directly.
Finally, FinRobot provides hands-on for both professional-grade analysts and laypersons to utilize powerful AI techniques for advanced financial analysis.
We open-source FinRobot at \url{https://github.com/AI4Finance-Foundation/FinRobot}.
\end{abstract}

\section{Introduction}
Financial analysis typically involves interpreting market trends, predicting economic outcomes, and providing investment strategies \cite{greenwald2004value,penman2010accounting,berman2013financial}. Financial analysis precedes financial decisions. Suppose we were to distinguish analysis performed in the equity space into two categories. Then, there is fundamental analysis \cite{abarbanell1997fundamental} that studies the companies and gives valuations, while technical analysis studies market action to forecast future price trends \cite{murphy1999technical}.

The analysis is grounded in data. As the digital revolution progresses, more data becomes accessible for analysis. Given the ever-growing volume and complexity of the data involved in the process, financial professionals increasingly resort to algorithmic and machine intelligence for data processing \cite{awotunde2021application}. Artificial intelligence (AI) \cite{cao2022ai} has revolutionized financial analysis by automating tasks typically performed by human analysts, such as sentiment analysis \cite{medhat2014sentiment,sohangir2018big,huang2023finbert,zhang2023fingptrag} and market prediction \cite{henrique2019literature,nabipour2020deep,kumar2022systematic,jiang2021applications}. Traditional AI models were initially limited to straightforward, single-task operations. However, advancements in LLMs \cite{brown2020language,wu2023bloomberggpt,yang2023fingpt} and computing power have significantly enhanced AI's capabilities. LLMs, with their extensive parameter space and diverse training datasets, are adept at assimilating global knowledge, understanding contexts, and executing complex logical reasoning. This expansion in capabilities not only allows managing multiple tasks simultaneously but also extends utility in the financial sector for more complex analysis and decision-making.

Building on the foundation of LLMs, AI agents have evolved to employ these models for an array of sophisticated functionalities, including planning, memory management, and tool usage, which facilitate intelligent and autonomous actions in real-world scenarios \cite{xi2023rise,gao2023large,wang2023survey,durante2024agent}. These capabilities allow AI agents to execute complex financial analyses and operations that traditionally required extensive human intervention and expertise. Recent applications of AI agents, such as FinAgent \cite{zhang2024finagent} for trading and FinMem \cite{yu2023finmem} for financial decision-making, underscore the growing reliance on these technologies to perform sophisticated financial operations. Despite their advances, several critical challenges remain unaddressed by these solutions:
\begin{itemize}[leftmargin=*]

\item \textbf{Transparency Improvement}: In what ways can AI Agent-driven financial analysis platforms improve transparency in their decision-making processes to build trust among users?

\item \textbf{Global Markets Adaptation}: How can AI agents effectively adapt to global stock markets' multilingual and multicultural nuances to ensure comprehensive and accurate market analysis?

\item \textbf{Model Diversity}: What are the limitations of relying solely on a single LLM architecture, such as GPT-4, for complex financial analysis, and how can diversifying model architectures enhance performance?

\item \textbf{Real-Time Data Processing}: How can AI agent platforms efficiently manage and route vast amounts of financial data to ensure timely and precise financial analysis?

\end{itemize}

Recognizing these challenges, this paper introduces FinRobot, an innovative open-source AI agent platform that harnesses multi-source LLMs for diverse financial tasks. FinRobot refines financial workflows from data processing to strategy implementation, broadening access to advanced analytical tools and enhancing scalability and transparency. By utilizing diverse LLM architectures, it tailors solutions to global market needs, ensuring precise adaptation and performance optimization. The platform's multi-source integration architecture facilitates the selection of accurate LLMs for specific tasks, and its real-time data processing pipeline assures timely financial analysis.

Our contributions are significant and address previous shortcomings in the field:
\begin{itemize}[leftmargin=*]

\item \textbf{Comprehensive Financial AI Agent Framework}: FinRobot provides a holistic framework for developing financial AI agents that can perform a wide range of tasks from market forecasting to financial document analysis. Notably, FinRobot is the first AI agent platform dedicated to financial analysis, marking a significant advancement in the application of AI in finance.

 \item \textbf{Multi-Source LLM Integration}: FinRobot introduces a novel Smart Scheduler mechanism that allows for the seamless integration of multi-source LLMs, enabling the platform to leverage the strengths of various state-of-the-art LLMs and select the most suitable ones for specific financial tasks. This adaptability is crucial for handling the complexities of global financial markets and multilingual data.

\item \textbf{Open-Source Platform for Financial AI}: By adopting an open-source approach, FinRobot fosters broader collaboration and accelerates innovation within the financial AI community. This openness not only promotes transparency but also allows for continuous enhancements and adaptations, which are essential in the rapidly evolving financial sector.
\end{itemize}

The remainder of this paper is organized as follows. Section 2 briefly reviews related work. In Section 3, we describe the four layers in FinRobot. In Section 4, we present the financial Chain-of-Thought (CoT) Prompting technique. In Section 5, we present the performance evaluation for two demo applications. Section 6 concludes this work and points out directions for future work.

\begin{figure*}
\centering
\includegraphics[scale = 0.33]{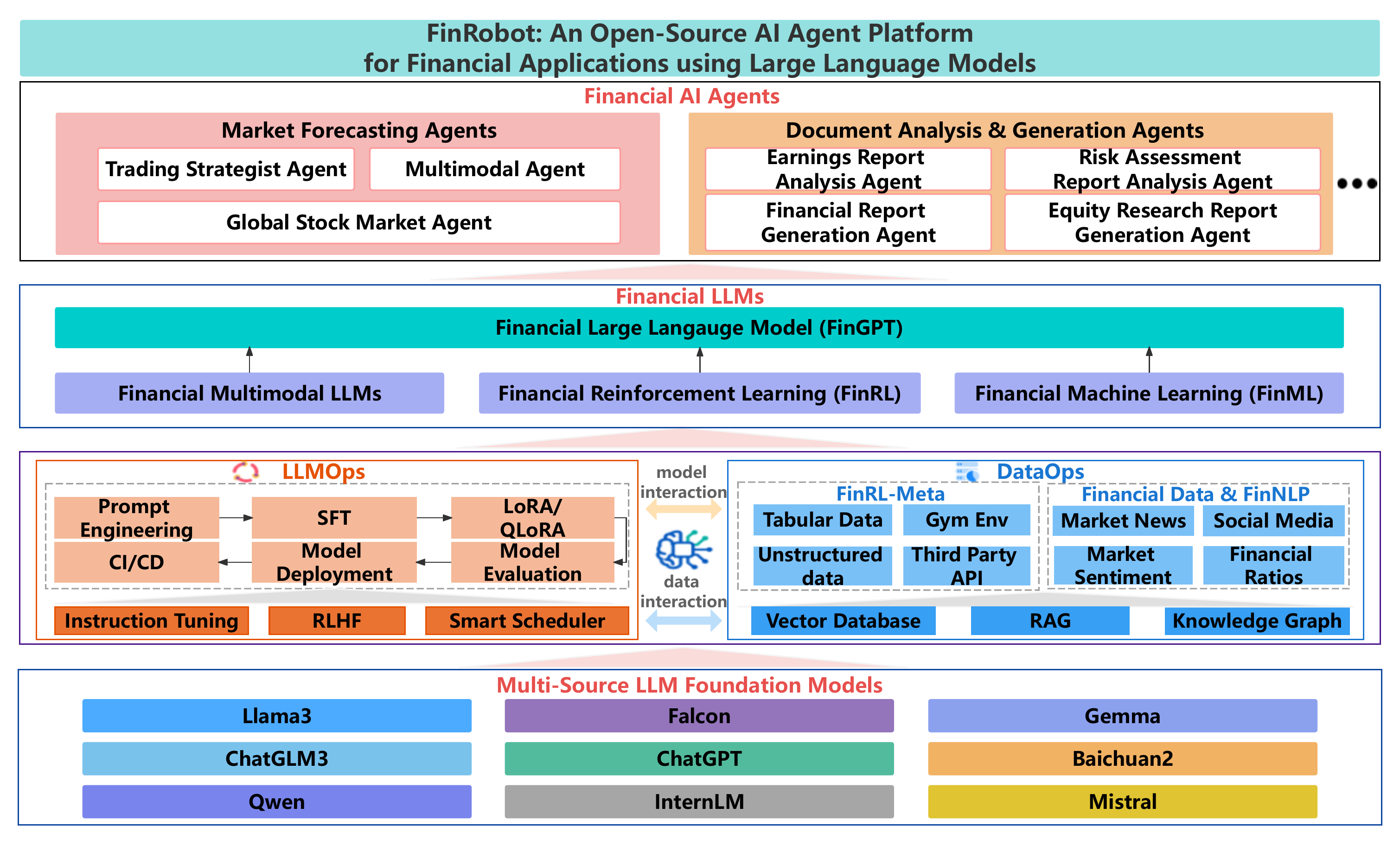}
\caption{Overall Framework of FinRobot.}
\label{fig:framework}
\vspace{-1mm}
\end{figure*}

\section{Related Work}

\subsection{LLM-Based AI Agents}
AI Agents, powered by large language models (LLMs), are intelligent entities capable of perceiving environments, making decisions, and executing actions independently. Recent advancements in AI have led to the proliferation of LLM-based AI agents across multiple domains, significantly impacting areas ranging from creative industries to technical fields \cite{sumers2023cognitive,park2023generative,zeng2023large}. For example, MusicAgent \cite{yu2023musicagent} enhances music composition, MedAgents \cite{tang2023medagents,thirunavukarasu2023large} improves medical diagnostics, Data Analytics Agent \cite{taskweaver} provides insights from big data, Programmer Agent \cite{rasheed2024codepori} automates coding tasks, and ResearchAgent generates research ideas \cite{baek2024researchagent}. These applications illustrate the versatility of LLMs and set the stage for their pivotal role in advancing financial technology.

\subsection{Financial AI Agents}
In the financial sector, AI agent-driven systems like FinAgent \cite{zhang2024finagent} and FinMem \cite{yu2023finmem} leverage LLMs to transform trading strategies by utilizing real-time market data for informed decision-making. However, the focus on metrics such as backtesting and individual stock returns often overshadows the need for robust, process-oriented approaches. Effective trading agents should not only perform well statistically but also enhance operational workflows, automating tasks from strategy execution to order placement with minimal human intervention. This shift from performance to process orientation can lead to more sustainable and adaptable financial technologies.

\subsection{Open-Source AI Agent Frameworks}
The development of open-source AI agent frameworks has been instrumental in democratizing access to advanced AI technologies. Prominent platforms and frameworks such as AutoGPT \cite{Significant_Gravitas_AutoGPT}, AutoGen \cite{wu2023autogen}, MetaGPT \cite{hong2023metagpt}, HuggingGPT \cite{shen2024hugginggpt}, ChatDev \cite{qian2023communicative}, Dify \cite{Charles2013} and Voyager \cite{wang2023voyager} facilitate collaborative enhancements and innovation by allowing a global community of developers to contribute and refine AI models. Given the growing need for sophisticated financial tools, the emergence of an open-source framework specialized for financial tasks is imminent.


\subsection{AI4Finance Foundation and Open-Source Culture}

The AI4Finance Foundation \footnote{\url{https://www.ai4finance.org}} plays a crucial role in shaping the open-source culture within financial technology, driven by its mission to promote standardized practices and develop open-source resources. This commitment is aimed at benefiting both researchers and industry professionals, fostering a collaborative environment where industry-specific knowledge meets and innovation thrives. By leading initiatives in applying AI technologies for financial services, AI4Finance not only accelerates technological advancements but also ensures these developments are accessible, transparent, and beneficial across the financial sector.



\section{Overview of FinRobot: An Open-Source Platform for Financial Tasks}

As delineated in Fig. \ref{fig:framework}, the overall framework of FinRobot is organized into four distinct layers, each designed to address specific aspects of financial AI processing and application:

\begin{itemize}

\item \textbf{Financial AI Agents Layer:} 
The Financial AI Agents Layer now includes Financial Chain-of-Thought (CoT) prompting, enhancing complex analysis and decision-making capacity. Market Forecasting Agents, Document Analysis Agents, and Trading Strategies Agents utilize CoT to dissect financial challenges into logical steps, aligning their advanced algorithms and domain expertise with the evolving dynamics of financial markets for precise, actionable insights.

\item \textbf{Financial LLMs Algorithms Layer:}  The Financial LLMs Algorithms Layer configures and utilizes specially tuned models tailored to specific domains and global market analysis. It employs FinGPT \cite{Wang2023FinGPTIT} alongside multi-source LLMs such as Llama series \cite{llama2} for the U.S. market and ChatGLM \cite{zeng2022glm} for the Chinese market, each optimized for regional specificities. The Falcon \cite{falcon40b} model excels in financial relationship analysis. Additionally, multimodal models integrate text with candlestick charts, while FinRL \cite{yang2020deep,finrl2020} optimizes tasks like portfolio allocation, and traditional machine learning methods \cite{Yang2018APM} refine stock selection. This approach ensures high precision in sensitive operations such as market forecasting and financial document analysis.

\item \textbf{LLMOps and DataOps Layers:} 
The LLMOps layer implements a multi-source integration strategy that selects the most suitable LLMs for specific financial tasks, utilizing a range of state-of-the-art models. Initially, general LLMs are deployed; if performance is suboptimal, the system dynamically switches to fine-tune LLMs to enhance effectiveness. This adaptive approach ensures tailored solutions for diverse financial scenarios, enhancing the platform’s overall performance. Concurrently, the DataOps layer manages real-time data processing \cite{liu2022finrl_meta,dynamic_datasets,2023finnlp}, which is crucial for rapid market responsiveness. This dual-layer configuration bolsters FinRobot’s ability to deliver timely and accurate financial insights under dynamic market conditions.

\item \textbf{Multi-source LLM Foundation Models Layer:} 
This foundational layer supports the plug-and-play functionality of various general and specialized LLMs. It forms the platform's backbone, ensuring that all models are up-to-date, optimized, and consistently aligned with the latest advancements in financial technologies and data standards.

\end{itemize}

\subsection{Financial AI Agents Layer}

\begin{figure}
\centering
\includegraphics[scale = 0.56]{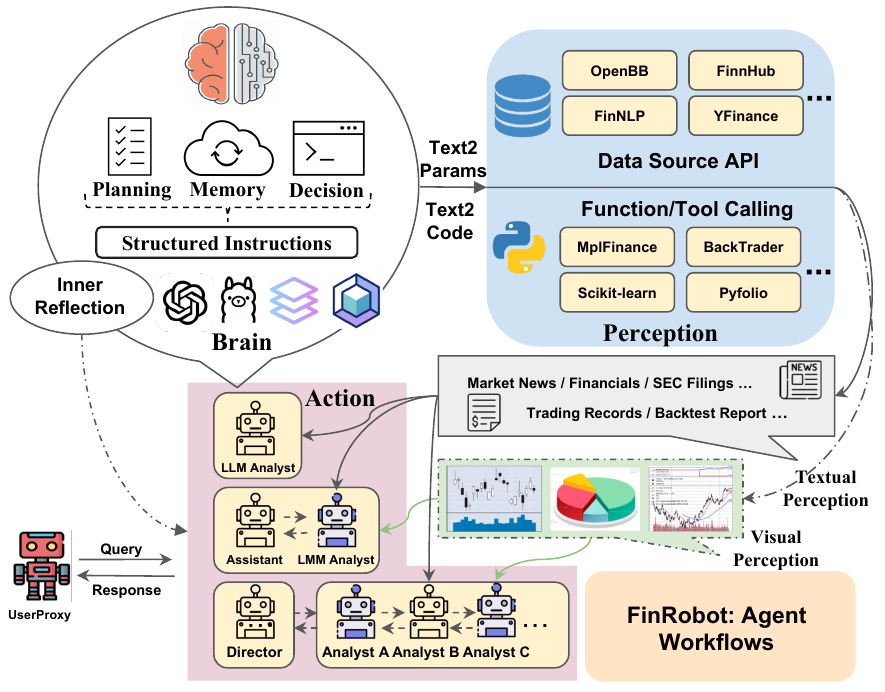}
\caption{Financial AI Agents Workflow Diagram}
\label{fig:agent_framework}
\vspace{-1mm}
\end{figure}

The Financial AI Agents Layer in FinRobot, as depicted in Fig. \ref{fig:agent_framework}, consists of domain-specific AI agents tailored to enhance financial analysis through advanced data perception, cognitive processing, and dynamic action execution:

\paragraph{Perception} This module captures and interprets multimodal financial data from market feeds, news, and economic indicators, using sophisticated techniques to structure the data for thorough analysis.

\paragraph{Brain} Acting as the core processing unit, this module perceives data from the Perception module with LLMs and utilizes Financial Chain-of-Thought (CoT) processes, as explained in Section 4, to generate structured instructions.

\paragraph{Action} This module executes instructions from the Brain module, applying tools to translate analytical insights into actionable outcomes. Actions include trading, portfolio adjustments, generating reports, or sending alerts, thereby actively influencing the financial environment.

\subsubsection{Multi-Agent Workflow}

Multi-agent workflow systems allow multiple agents, each with unique roles and responsibilities, to tackle intricate tasks in a cooperative manner. The deployment of a multi-agent workflow system is beneficial when navigating complex financial datasets and ensuring a high standard of analytical accuracy and depth. This system incorporates multiple specialized roles that work in concert to process, analyze, and derive actionable insights from diverse financial data sources.

\paragraph{Director} As the strategic leader of the project, the Director in a financial analysis context oversees all aspects of the workflow. This role entails prioritizing financial tasks, allocating resources, and orchestrating team efforts to optimize analysis timelines and outcomes. 

\paragraph{Assistant} The Assistant in financial analysis focuses on initial data management tasks such as collecting, processing, and conducting preliminary analyses of financial data and facilitates communication between agents. This role supports more specialized analysts by preparing datasets, maintaining databases, and executing basic financial computations and visualizations. This foundational work enables other analysts to focus on more complex and interpretive aspects of financial analysis.

\paragraph{LLM Analyst} The LLM Analyst employs advanced computational techniques to analyze financial texts, such as reports, filings, and news articles, using LLMs. This role is essential for extracting detailed insights, conducting sentiment analysis, and forecasting market trends from qualitative data. By integrating these insights with quantitative data, the LLM Analyst provides a comprehensive assessment of financial health and market conditions, enhancing the decision-making process.

\paragraph{Financial Analysts} Operating under the LLM Analyst's guidance, Financial Analysts engage in detailed quantitative data analysis across various sectors like portfolio management, risk assessment, and market analysis. Utilizing statistical tools and financial models, they interpret data, evaluate investment opportunities, and formulate risk mitigation strategies. Their expert analysis contributes significantly to the development of sound financial strategies and informed recommendations.

The collaboration among these roles within the multi-agent workflow enables a holistic approach to financial analysis, addressing macroeconomic conditions and specific financial metrics. This structured yet flexible approach ensures thorough exploration and interpretation of financial data, leading to informed strategic decision-making.

\subsubsection{Utilizing LLMs for Tool Usage}

\paragraph{API Interaction through Text2Params} This method translates natural language queries into API requests \cite{wang2024tools}, effectively combining the capabilities of generating function calls and forming API calls. In this approach, the LLM first parses the text to identify and extract key parameters. These parameters are then utilized to dynamically generate function calls or compile directly into API requests that interact with financial software or databases. This method is especially useful for executing predefined operations and fetching or manipulating financial data from external services.

\paragraph{Code Compilation through Text2Code} For more complex financial tasks that require dynamic solution generation, LLMs can employ text2code techniques to write and compile code on the fly. This capability is essential for developing custom algorithms based on user queries' unique market conditions or financial scenarios.

\subsection{Financial LLMs Algorithms Layer:}
This layer encompasses advanced AI algorithms specifically designed to address various needs within the financial sector, enhancing the platform's capabilities across a range of financial applications.

\subsubsection{Financial Large Language Models (FinGPT)}
FinGPT \cite{yang2023fingpt}, a domain-specific LLM, is meticulously engineered to elevate natural language understanding specifically within the financial context. These models adeptly analyze and interpret financial narratives, extracting critical data from complex documents such as annual reports and real-time financial news, thereby supporting enhanced decision-making processes.

The FinGPT models are mainly developed by supervised finetuning with the financial in-domain paired ``instruction-response'' data on the open-source large language models. This process is achieved by minimizing the following negative log-likelihood,
\begin{equation}
    \mathcal{L}_{\text{CausalLM}} = - \sum_{t=1}^{T} \log P(w_t | w_1, w_2, \ldots, w_{t-1}; \theta),
\end{equation}
where $T$ is the length of the input sequence, $w_i$ represents the $i$-th token in the sequence, $\theta$ denotes the model parameters, and $P(w_i | w_1, w_2, \ldots, w_{i-1}; \theta)$ represents the conditional probability of predicting the target token $x_i$ given the preceding tokens $x_1, \ldots, x_{i-1}$. By optimizing this objective function, the model learns to maximize the probability of generating the expected response given the finance task instruction.

\subsubsection{Financial Reinforcement Learning (FinRL)}
FinRL \cite{yang2020deep} optimizes trading strategies using ensemble deep RL algorithms to analyze historical and real-time market data. This dynamic adaptation helps maximize financial returns while minimizing risks, making stock portfolio allocation an ideal application for this methodology.

In FinRL, stock portfolio allocation is modeled as a Markov Decision Process (MDP). At any time $t$, an agent in state $s_t \in \mathcal{S}$ selects an action $a_t \in \mathcal{A}$ based on policy $\pi_{\theta}(s_t)$. This action leads to a new state $s_{t+1}$ and a reward $r(s_t, a_t, s_{t+1})$. The goal is to optimize this policy:

\begin{equation}
\begin{split}
    \pi^*_{\theta} &= \underset{\theta}{\text{argmax}}~ J(\pi_\theta), \\
    \text{where} \quad J(\pi_\theta) &= \mathbb{E}\left[\sum^{T}_{t=0} \gamma^t r(s_t, a_t, s_{t+1})\right],
\end{split}
\end{equation}
where $\gamma \in (0,1]$ is the discount factor.

\subsubsection{Financial Machine Learning (FinML)}
FinML \cite{Yang2018APM} leverages diverse machine learning techniques, from regression to advanced neural networks, to boost predictive analytics in finance. These algorithms play a crucial role in forecasting market trends, consumer behavior, credit risks, and other pivotal financial indicators, thus facilitating informed decision-making. A key measure used by FinML is the log-return, calculated as follows:

\begin{equation}
r_{T+f,i} = \log\left(\frac{S_{T+f,i}}{S_{T,i}}\right), \quad i = 1, \ldots, n_T,
\end{equation}
where $r$ denotes the log-return, $S$ is the stock price, $n_T$ represents the number of companies monitored at time $T$, and $f$ is the forecast horizon. This measure is essential for evaluating investment performance and shaping financial strategies.

\subsubsection{Financial Multimodal LLMs}
Financial documents frequently incorporate diverse data types beyond textual content, such as graphs \cite{bhatia2024fintral} and tables \cite{wang2023docllm}. These additional modalities offer rich, complementary insights, significantly enhancing the depth of analysis possible. To effectively integrate and leverage these varied data types, we have developed Financial Multimodal LLMs, which is specifically engineered to process and synthesize information from multiple modalities, thereby providing a comprehensive and nuanced understanding of financial documents.

The mathematical representation of the integration process in our Financial Multimodal LLM is as follows:
\begin{equation}
    F(x_t, x_g, x_h) = L(T(x_t), G(x_v), H(x_l))
\end{equation}

where $F(x_t, x_g, x_h)$ denotes the output of the model. Here, $x_t$, $x_g$ and $x_h$ represent the inputs of textual, graphical, and tabular data, respectively. The functions $T$, $G$ and $H$ transform these inputs into a unified embedding space. The LLM $L$ then synthesizes these embeddings to produce a coherent and reliable output, enhancing both the accuracy and reliability of the financial analyses.

\subsection{LLMOps Layer}

\begin{figure}[t]
\centering
\includegraphics[scale = 0.45]{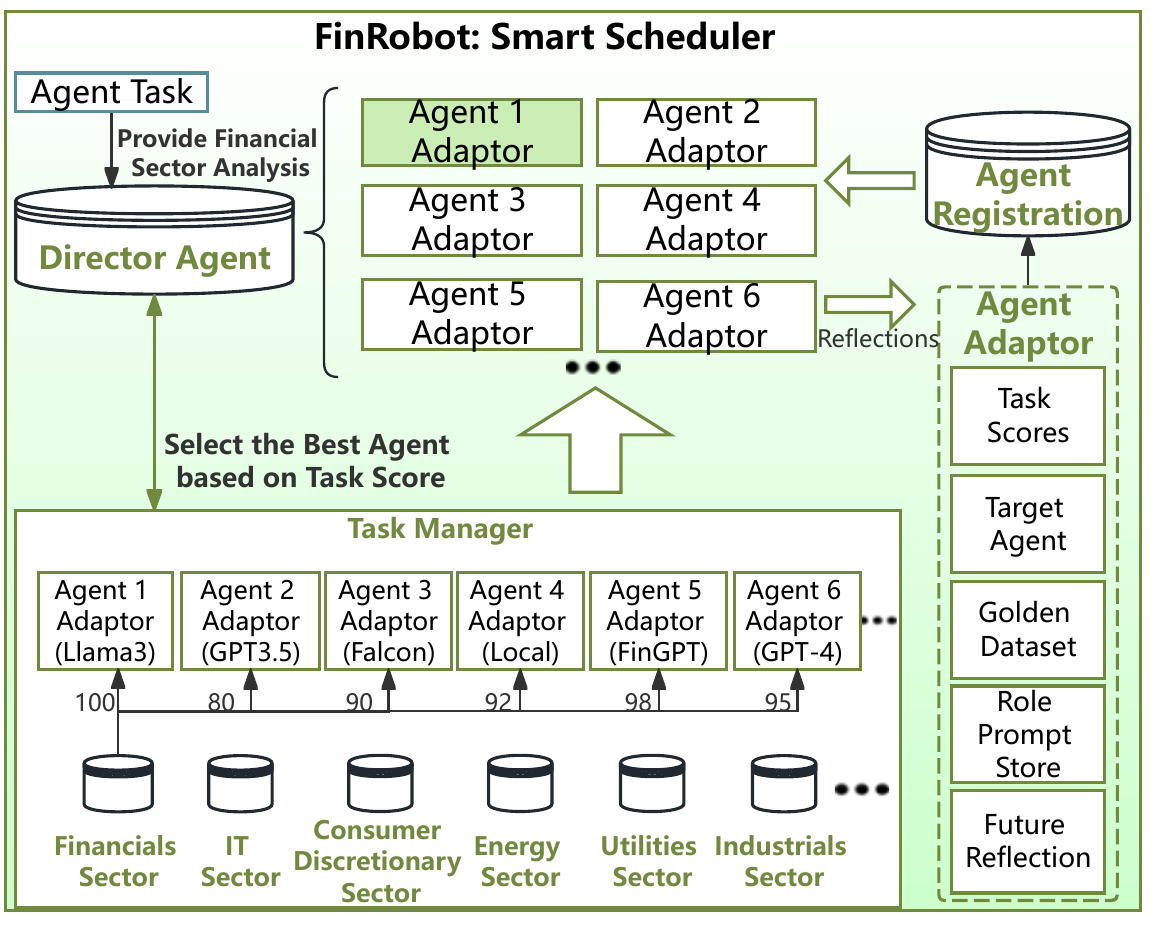}
\caption{Smart Scheduler Structural Layout in LLMOps Layer}
\label{fig:scheduler}
\vspace{-1mm}
\end{figure}

The LLMOps Layer is designed for high modularity and pluggability, accommodating rapid integration and dynamic swapping of LLMs in response to evolving technological advancements and financial market demands. This layer facilitates seamless model integration and includes mechanisms for rigorous evaluation and selection of the most suitable models for specific financial tasks. These capabilities are critical for maintaining operational efficiency and ensuring the adaptability of solutions to diverse financial scenarios.

\subsubsection{Smart Scheduler}
The Smart Scheduler, depicted in Figure \ref{fig:scheduler}, is central to ensuring model diversity and optimizing the integration and selection of the most appropriate LLM for each task. 

\subsubsection{Smart Scheduler Architecture}
The Smart Scheduler optimizes task distribution among agents and comprises:
\begin{itemize}
    \item \textbf{Director Agent}: This component orchestrates the task assignment process, ensuring that tasks are allocated to agents based on their performance metrics and suitability for specific tasks.
    \item \textbf{Agent Registration}: Manages the registration and tracks the availability of agents within the system, facilitating an efficient task allocation process.
    \item \textbf{Agent Adaptor}: Tailors agent functionalities to specific tasks, enhancing their performance and integration within the overall system.
    \item \textbf{Task Manager}: Manages and stores different general and fine-tuned LLMs-based agents tailored for various financial tasks, updated periodically to ensure relevance and efficacy.
\end{itemize}

\subsubsection{Smart Scheduler Initialization Process}
The initialization process is streamlined to establish a robust foundation for agent operations effectively:
\begin{itemize}
    \item \textbf{Golden Dataset Creation}: Populates with industry-specific data
    \item \textbf{Prompt Store Population}: Fills with custom prompts for various agents.
    \item \textbf{Task Score Filling}: Invokes adaptors to evaluate responses against best answers, storing scores in the Task Scores database.
\end{itemize}

\subsubsection{Smart Scheduler In Action Process}
The operational phase follows this sequence:
\begin{enumerate}
    \item \textbf{Task Initiation}: A user initiates a task.
    \item \textbf{Director Agent's Role}: Evaluates task inputs, ranking agents by performance and relevance.
    \item \textbf{Agent Selection and Task Routing}: Routes tasks to the highest-ranked agent
    \item \textbf{Workflow Progression and Self-Evaluation}: Post-task completion, the agent performs a self-assessment, the results of which are stored in the Future Reflection storage.
    \item \textbf{Workflow Completion and Evaluation}: At the end of the workflow, the agent evaluates the results and provides feedback for continuous improvement.
\end{enumerate}

\subsubsection{Scoring Metrics}
The scoring process within the Smart Scheduler involves several critical steps to evaluate the performance of different LLMs for task-specific applications:

\begin{enumerate}
\item \textbf{Data Collection}: Collect performance data of various LLMs across multiple evaluation tasks.
\item \textbf{Normalization}: Normalize the results for each evaluation task to scale between 0 and 1.
\item \textbf{Weight Assignment}: Assign weights to different evaluation dimensions based on industry standards or expert opinions.
\item \textbf{Calculation of Composite Score}: Multiply the normalized scores by their respective weights and sum them to derive the task score.
\item \textbf{Results Analysis}: Analyze the scores to rank and assess the LLMs, providing performance comparisons and recommendations for selection.
\end{enumerate}
This enhanced structure not only boosts the operational efficiency of financial AI agents but also supports scalable, dynamic management of diverse financial tasks, significantly improving the system's capability to manage complex, multi-agent scenarios in real-time financial environments.

\subsection{DataOps Layer}
The DataOps Layer manages the extensive and varied datasets necessary for financial analysis, ranging from public tabular data to proprietary market sentiments. This layer is crucial in ensuring that all data fed into the AI processing pipelines is of high quality and representative of the current market conditions. Effective data management is foundational for the accurate and reliable performance of AI models, and as such, the DataOps Layer employs advanced data handling and processing techniques to prepare and deliver these inputs. By optimizing data accessibility and quality, the DataOps Layer supports the overall efficacy of the FinRobot platform, enabling sophisticated and precise financial decision-making based on solid data-driven insights.

\subsubsection{Retrieval Augmented Generation} Retrieval-augmented generation (RAG) is a pivotal technique incorporated within FinGPT \cite{zhang2023fingptrag,zhang2023instructfingpt}, as it seamlessly combines the strength of both context retrieval mechanisms and LLMs to optimize language generation tasks. 

The LLMOps and DataOps layers form the backbone of FinRobot’s operational capabilities, ensuring that the platform remains at the forefront of technological innovation while providing reliable and effective financial AI services.

\subsection{Multi-source LLM Foundation Models Layer}
The Multi-source LLM Foundation Models Layer equips FinRobot with advanced capabilities to manage and integrate a diverse array of Large Language Models (LLMs), essential for adapting to the dynamic demands of global financial markets. Key features of this layer include:

\begin{itemize}
\item \textbf{Plug-and-Play Functionality:} Enables seamless integration and updating of both general and specialized LLMs, ensuring the platform remains adaptable and up-to-date with financial technology advancements.
\item \textbf{Model Diversity and Evaluation:} Incorporates LLMs with parameters ranging from 7 billion to 72 billion, each rigorously evaluated for effectiveness in specific financial tasks, allowing for optimal model selection based on performance metrics such as accuracy and adaptability.
\item \textbf{Global Market Compatibility:} Supports multilingual model integration, enhancing the platform’s ability to analyze and process diverse financial data, which is crucial for global market operations.
\end{itemize}

This streamlined version succinctly outlines the layer's functionality and strategic role in maintaining FinRobot’s leadership in AI-driven financial analysis.

\section{Financial Chain-of-Thought (CoT) Prompting}

\subsection{Introduction}

The chain-of-thought prompting technique \cite{wei2022chain,xia2024beyond} structures prompts to foster a step-by-step reasoning process within AI models, akin to human problem-solving strategies. This method significantly enhances performance on complex reasoning tasks such as mathematics and commonsense reasoning by encouraging models to articulate intermediate reasoning steps, leading to a final answer. This method improves accuracy and enhances the interpretability and transparency of the decision-making process.

\subsection{Concept of Financial CoT}

Financial CoT Prompting adapts the CoT technique for AI-driven financial analysis, integrating advanced cognitive processing techniques to refine decision-making capabilities. It stimulates the thought process of financial professionals to tackle complex financial issues by guiding AI models through a logical, sequential reasoning process. This approach breaks down complex financial scenarios into smaller, manageable components, analyzes each component, and synthesizes the findings to form conclusions or recommendations. As noted by \cite{kim2024financial}, this method mimics human-like reasoning in the analysis chain and is particularly valuable for tasks requiring deep analytical thought, such as valuation, investment strategy formulation, market trend analysis, and risk assessment.

\subsection{Financial CoT Implementation}

Financial Chain-of-Thought (CoT) Prompting marks a transformative shift in security analysis, enhancing traditional analytical tools across various domains by leveraging the capabilities of LLMs such as GPT-4. Detailed case studies, such as those conducted by \cite{kim2024financial}, highlight this approach's practical applications and benefits.

\begin{itemize}
    \item \textbf{Financial Analysis}: FinRobot transcends basic data enumeration and ratio computation. By employing LLMs for financial statement analysis, it conducts comparative analyses across industry competitors and against the company's historical performance. This comprehensive approach provides profound insights into a company's business landscape and the logical underpinnings of its financial metrics. Specifically, it normalizes financial ratios to identify anomalies, enhancing the accuracy of financial insights beyond human capabilities.
    
    \item \textbf{Business-Specific Analysis}: Leveraging Retrieval-Augmented Generation (RAG), FinRobot revolutionizes the process of gathering data on a company's products and services, complementing the information available in financial reports. By tapping into the vast resources of the web, FinRobot enriches its understanding of a company's offerings, enabling a comprehensive analysis of product lines, trends across channels and regions, cost structures, supply chain dynamics, and R\&D conversions. 
    
    \item \textbf{Market Analysis}: FinRobot abstracts real-world financial contexts, utilizing a combination of financial ratios, market data (including stock price trends and candlestick charts), market news sentiment, and alternative data sourced online. By employing LLMs, FinRobot simulates market participants' decision-making processes in response to contextual changes. This allows for a comprehensive evaluation of a company's stock price trajectory and valuation. FinRobot provides nuanced investment assessments, offering strategic guidance on optimal investment timing and suitable financial instruments, ensuring robust, well-informed investment strategies considering various horizons, structures, and risk thresholds. 
    
    \item \textbf{Valuation Analysis}: FinRobot integrates its comprehensive analysis of financial ratios, market data (including stock price trends and candlestick charts), market news sentiment, and alternative data collected online. FinRobot employs LLMs to evaluate a company's stock price trajectory and valuation using the insights derived from these diverse sources. This approach enables FinRobot to provide nuanced investment assessments across various horizons, structures, and risk thresholds, offering strategic guidance on optimal investment timing and suitable financial instruments, thereby ensuring well-informed and robust investment strategies. 
\end{itemize}

\subsection{Advantages of Financial CoT Prompting}

Despite their merits, a readily available tool that can intelligently perform industry- and company-specific analytics still needs to exist in a way that parallels the work of an investment professional. Existing data providers such as Bloomberg, FactSet, CapitalIQ, and Refinitiv Eikon rely on rudimentary, generic templates for data extraction, inadvertently overlooking key operational metrics and qualitative information intricately linked with the business essence. Manual analysis can be time-consuming and occasionally repetitive while existing automated tools can only sometimes extract raw data from unstructured sources like annual reports. Although these platforms abound with vast amounts of information, extracting essential information and its subsequent analysis still require the expertise of investment professionals. 

With CoT Prompting, FinRobot establishes a unique niche among AI Agent Platforms by emulating human cognitive processes in financial analysis, transcending traditional reliance on mere numerical computations. Our methodology leverages LLMs to deconstruct and analyze the methodologies utilized by financial professionals when scrutinizing business data. This approach unveils a practical analysis intrinsically dependent on a comprehensive understanding of a company's business fundamentals, as reflected through detailed accounting data, tailored industry-specific metrics, and qualitative information.

This approach overcomes the above limitations because:
(1) It does not rely on a rigid template due to the generative nature of LLMs.
(2) It avoids misclassification because error-checking is embedded within layers of prompting.
(3) It simplifies human labor by codifying investment logic within its reasoning.
(4)  It can extract relevant historical numbers and qualitative information from complex and unstructured data formats such as earnings call transcripts.
(5) Given its multi-layered structure, it provides stronger grounds for explaining the sources and derivations of each recorded and derived value.
(6)  It is highly adaptable and evolving as LLMs may be improved through supervised fine-tuning and instruction-tuning.

\subsection{Market Simulation: Transcending the Analyst Mimicking}

To expand the concept of market simulation and explore its possibilities, one could view the GPT's ability to mimic human analytical processes as a gateway to simulating broader market dynamics. Market simulation transcends mere numerical analysis by incorporating human-like reasoning processes, embodying multiple personas and scenarios within a financial context. This simulation could encompass a variety of market participants whose decisions are influenced by an evolving set of data inputs—ranging from financial statements, reports, market data, and economic indicators to global news.

The core of this approach involves constructing a detailed simulation environment where these generative agents, representing different market players, interact \cite{park2023generative}. Each agent's decision-making process would be guided by data-driven prompts, mimicking the cognitive steps a human analyst might take. This includes interpreting complex market signals and reacting to new information in real time, which advanced machine learning models and reinforcement learning techniques could facilitate.

By leveraging LLMs, one could theoretically encode these decision-making frameworks into the agents, enabling them to perform tasks that require a deep understanding of the market's quantitative and qualitative aspects. This would simulate individual trading decisions and the broader market dynamics resulting from multiple decision-makers interactions. The potential to use this technology to create a virtual market environment where hypothetical scenarios could be tested and analyzed could provide invaluable insights into market behavior and strategy development.

A detailed framework for such a simulation could involve:
(1) Defining the roles and behaviors of different market participants (e.g., institutional investors, retail investors, market makers).
(2) Implementing a variety of data inputs and decision-making models for each participant.
(3) Simulating market interactions over different time horizons and under varying conditions to assess potential outcomes and strategies.
(4) Utilizing reinforcement learning to refine the agents' strategies based on observed performance in the simulation.

As outlined by \cite{park2023generative}, this concept presents a promising avenue for future research and development in financial market analysis and strategy optimization.

\section{Demo Applications and Hands-on Tutorial}

\subsection{Application I: Market Forecaster}
Market Forecaster is a set of AI Agents designed to synthesize recent market news and financial data, delivering comprehensive insights into a company's latest achievements and potential concerns, along with predictions for stock price movements. As a junior robo-advisor, Market Forecaster embodies a substantial leap forward in AI-driven financial advisory.

\begin{figure}[ht!]
\centering
\begin{tcolorbox}[
  enhanced,
  colback=white,
  boxrule=0.5pt,
  arc=4pt,
  left=6pt,
  right=6pt,
  top=6pt,
  bottom=6pt
]
\small 
\textbf{Instruction:} You are an experienced stock market analyst. Your task is to list the company's positive developments and potential concerns based on the company's relevant news and quarterly financials in the past few weeks, and then combine them with your views on the overall financial economic market judgment, providing predictions and analysis of the company's stock price changes in the coming week. Your answer format should be as follows: \\

{[Positive development]:}\\
1. ...

{[Potential concerns]:}\\
1. ...

{[Forecast and Analysis]:}\\
...\\

\textbf{Information:}\\
a. Company Introduction\\
b. Stock Price Changes\\
c. Recent News Information\\
d. Recent Basic Financials\\

\textbf{Instruction:} 
Based on all the information before 2024-04-19, let's first analyze the positive developments and potential concerns for AAPL. Come up with 2-4 most important factors respectively and keep them concise. Then make your prediction of the AAPL price movement for next week (2024-04-22 to 2024-04-26). Provide a summary analysis to support your prediction.

\end{tcolorbox}
\vspace{-2mm}
\caption{Market Forecaster Prompt Template}
\label{fig:forecaster_prompt_template}
\vspace{-3mm}
\end{figure}

\begin{figure}[ht!]
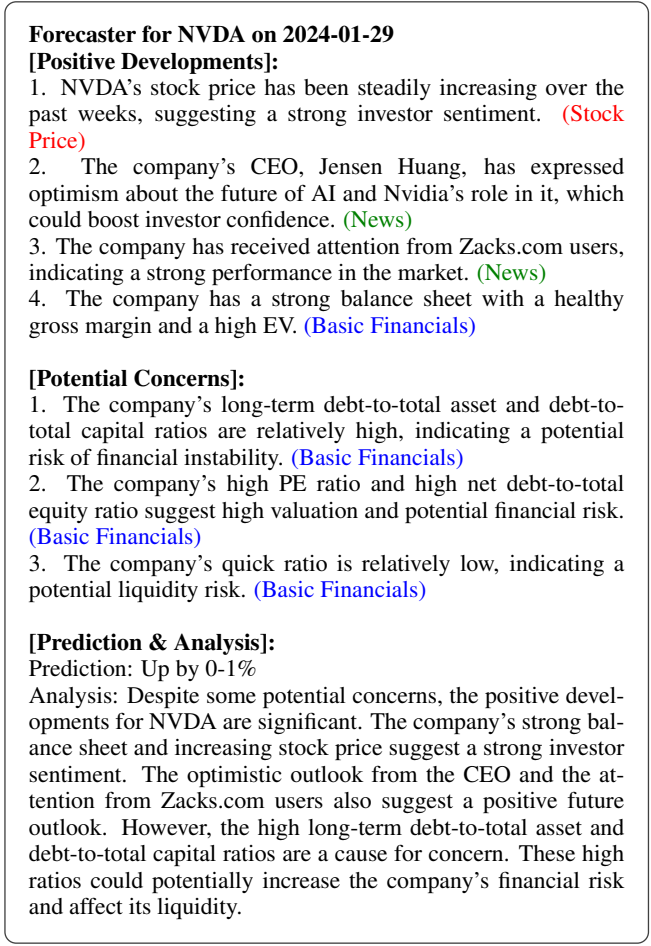

\centering
\begin{tcolorbox}[
  enhanced,
  colback=white,
  boxrule=0.5pt,
  arc=4pt,
  left=6pt,
  right=6pt,
  top=6pt,
  bottom=6pt
]
\small 
\textbf{Forecaster for NVDA on 2024-01-29}

\textbf{[Positive Developments]:}\\
1. NVDA's stock price has been steadily increasing over the past weeks, suggesting a strong investor sentiment. \textcolor{stockPriceColor}{(Stock Price)}\\
2. The company's CEO, Jensen Huang, has expressed optimism about the future of AI and Nvidia's role in it, which could boost investor confidence. \textcolor{newsColor}{(News)}\\
3. The company has received attention from Zacks.com users, indicating a strong performance in the market. \textcolor{newsColor}{(News)}\\
4. The company has a strong balance sheet with a healthy gross margin and a high EV. \textcolor{financeColor}{(Basic Financials)}\\

\textbf{[Potential Concerns]:}\\
1. The company's long-term debt-to-total asset and debt-to-total capital ratios are relatively high, indicating a potential risk of financial instability. \textcolor{financeColor}{(Basic Financials)}\\
2. The company's high PE ratio and high net debt-to-total equity ratio suggest high valuation and potential financial risk. \textcolor{financeColor}{(Basic Financials)}\\
3. The company's quick ratio is relatively low, indicating a potential liquidity risk. \textcolor{financeColor}{(Basic Financials)}\\

\textbf{[Prediction \& Analysis]:}\\
Prediction: Up by 0-1\% \\
Analysis: Despite some potential concerns, the positive developments for NVDA are significant. The company's strong balance sheet and increasing stock price suggest a strong investor sentiment. The optimistic outlook from the CEO and the attention from Zacks.com users also suggest a positive future outlook. However, the high long-term debt-to-total asset and debt-to-total capital ratios are a cause for concern. These high ratios could potentially increase the company's financial risk and affect its liquidity.
\end{tcolorbox}
\vspace{-3mm}
\caption{Market Forecaster Result of Nvidia}
\label{fig:forecaster_ans1}
\vspace{-3mm}
\end{figure}

\begin{figure}[ht!]
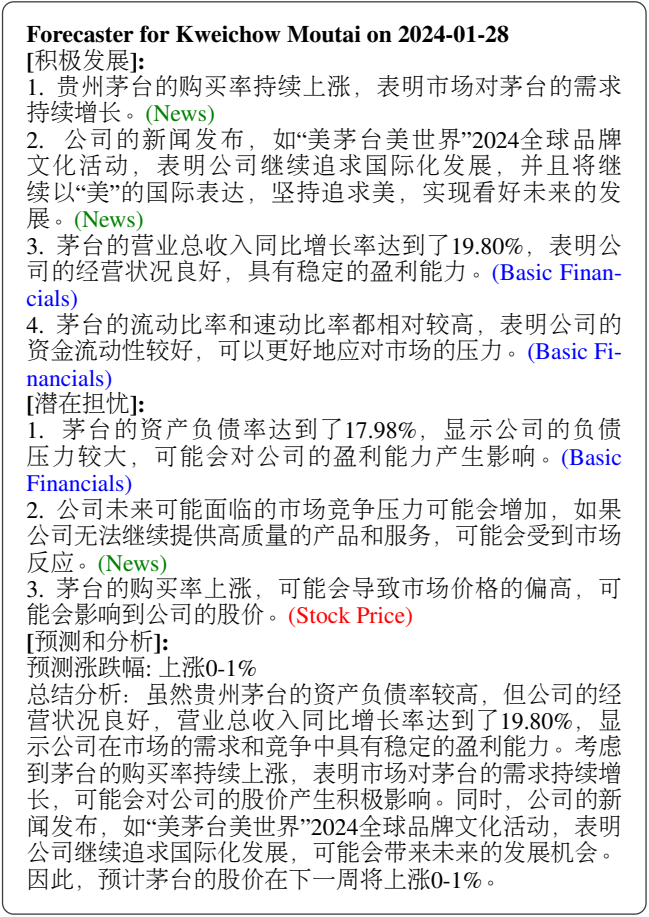

\centering
\begin{tcolorbox}[
  enhanced,
  colback=white,
  boxrule=0.5pt,
  arc=4pt,
  left=6pt,
  right=6pt,
  top=6pt,
  bottom=6pt
]
\small
\begin{CJK}{UTF8}{gbsn} 
\textbf{Forecaster for Kweichow Moutai on 2024-01-28}

\textbf{[积极发展]:}\\
1. 贵州茅台的购买率持续上涨，表明市场对茅台的需求持续增长。\textcolor{newsColor}{(News)}\\
2. 公司的新闻发布，如“美茅台美世界”2024全球品牌文化活动，表明公司继续追求国际化发展，并且将继续以“美”的国际表达，坚持追求美，实现看好未来的发展。\textcolor{newsColor}{(News)}\\
3. 茅台的营业总收入同比增长率达到了19.80\%，表明公司的经营状况良好，具有稳定的盈利能力。\textcolor{financeColor}{(Basic Financials)}\\
4. 茅台的流动比率和速动比率都相对较高，表明公司的资金流动性较好，可以更好地应对市场的压力。\textcolor{financeColor}{(Basic Financials)}
\end{CJK} 

\begin{CJK}{UTF8}{gbsn} 
\textbf{[潜在担忧]:}\\
1. 茅台的资产负债率达到了17.98\%，显示公司的负债压力较大，可能会对公司的盈利能力产生影响。\textcolor{financeColor}{(Basic Financials)}\\
2. 公司未来可能面临的市场竞争压力可能会增加，如果公司无法继续提供高质量的产品和服务，可能会受到市场反应。\textcolor{newsColor}{(News)}\\
3. 茅台的购买率上涨，可能会导致市场价格的偏高，可能会影响到公司的股价。\textcolor{stockPriceColor}{(Stock Price)}
\end{CJK} 

\begin{CJK}{UTF8}{gbsn} 
\textbf{[预测和分析]:}\\
预测涨跌幅: 上涨0-1\% \\
总结分析：虽然贵州茅台的资产负债率较高，但公司的经营状况良好，营业总收入同比增长率达到了19.80\%，显示公司在市场的需求和竞争中具有稳定的盈利能力。考虑到茅台的购买率持续上涨，表明市场对茅台的需求持续增长，可能会对公司的股价产生积极影响。同时，公司的新闻发布，如“美茅台美世界”2024全球品牌文化活动，表明公司继续追求国际化发展，可能会带来未来的发展机会。因此，预计茅台的股价在下一周将上涨0-1\%。
\end{CJK} 
\end{tcolorbox}
\vspace{-3mm}
\caption{Market Forecaster Result of Kweichow Moutai}
\label{fig:forecaster_ans2}
\vspace{-3mm}
\end{figure}

\subsubsection{Data}
Market Forecaster stands at the crossroads of global markets, leveraging diverse data sources to gather information and make forecasting decisions. In detail, Market Forecaster gathers data regarding multifaceted company information such as recent news, latest basic financials, and stock prices for the target markets, including the US stock market, Chinese stock market, Crypto market, and other possible extensions. This section will give a solid overview of the Market Forecaster targeted at the US stock market and Chinese stock market with data sources Finnhub and EastMoney.

\subsubsection{Model}
This AI Agent is powered by fine-tuned FinGPT-Forecasters using Llama-2-7b-chat-hf  with LoRA, leveraging data from the latest year's DOW 30 for the US Market and SSE 50 for the Chinese Market to ensure precise forecasts for these major stocks and demonstrating robust generalization abilities across various stock symbols. Moreover, this paper follows the Financial Multi-task Instruction Tuning paradigm\cite{Wang2023FinGPTIT,yu2023harnessing} to align the power of the base model with the specific functions of Market Forecaster.

\subsubsection{Prompts}
Aligning with the multi-task instruction tuning framework, the Market Forecaster follows a sophisticated prompt format. Specifically, the Market Forecaster gathers data of multifaceted company information, then it performs prompt engineering to format the instructive prompt with the structure of ‘Task Instruction {\&} Company Information (Company Overview + Recent Stock Prices + Recent News + Latest Basic Financials)’. Figure 4 shows an example of the Market Forecaster prompt template.

\subsubsection{Sample Forecastings}
The sample answers provided for two prominent stocks in the US and Chinese markets, namely Nvidia and Kweichow Moutai, as depicted in Figures 5 and 6, respectively, illustrate the Market Forecaster's adeptness in synthesizing and offering valuable insights gleaned from diverse information sources. Additionally, the Market Forecaster furnishes recommendations regarding the future trajectory of the stocks, underscoring its capability to provide actionable guidance based on the analyzed data.

\subsection{Application II: Document Analysis \& Generation}

The Document Analysis \& Generation application represents a groundbreaking use of AI Agents and Large Language Models (LLMs) in the realm of financial document management and report creation. This application harnesses the power of AI to perform deep analysis of financial documents and generate detailed, insightful reports automatically.

\subsubsection{Document Analysis}

AI Agents, integrated with advanced LLMs, are employed to sift through extensive financial documents such as annual reports, SEC filings, and earnings call transcripts. These agents are capable of extracting critical information, identifying key financial indicators, and highlighting trends and discrepancies that may require closer inspection. The ability of these AI Agents to understand and process complex financial terminology and context turns vast amounts of unstructured data into structured, actionable insights.

\subsubsection{Report Generation}

Following the analysis phase, the same agent is utilized to generate comprehensive financial reports. Leveraging the capabilities of LLMs, the system produces coherent, articulate, and detailed documents that cover various aspects of financial analysis, including performance evaluations, market comparisons, and forward-looking financial forecasts. Each report is crafted to maintain a professional tone and format, mirroring the quality and depth expected of top-tier financial analysts. 

For practical insights, equity research reports generated by FinRobot are included in the appendix, showcasing the application and effectiveness of our platform.






\section{Conclusion}
FinRobot revolutionizes financial analysis by integrating multi-source Large Language Models (LLMs) in an open-source platform that enhances accessibility, efficiency, and transparency in financial operations. This innovative platform addresses the complexities of global markets with a multi-layered architecture that supports real-time data processing and diverse model integration, making sophisticated financial tools available to a wider audience. By fostering collaboration within the financial AI community, FinRobot not only accelerates innovation but also sets new standards for the application of AI in finance, promising to significantly improve strategic decision-making across the sector.

\textbf{Future Work}
As FinRobot continues to evolve, we plan to expand its applications to include more sophisticated tasks such as portfolio allocation and comprehensive risk assessment, enhancing its utility in the financial sector. Additionally, we aim to broaden the platform's reach by extending its capabilities to more global markets. This expansion will not only diversify the potential use cases of FinRobot but also improve its adaptability to different economic environments, driving innovation and accessibility in AI-driven financial analysis across the world.

\bibliographystyle{named}
\bibliography{ref}
\textbf{Disclaimer: We are sharing codes for academic purposes under the MIT education license. They should not be construed as financial counsel or recommendations for live trading. It is imperative to exercise caution and consult with qualified financial professionals prior to any trading or investment actions.}

\newpage

\appendix


\begin{figure*}
\centering
\includegraphics[scale = 0.77]{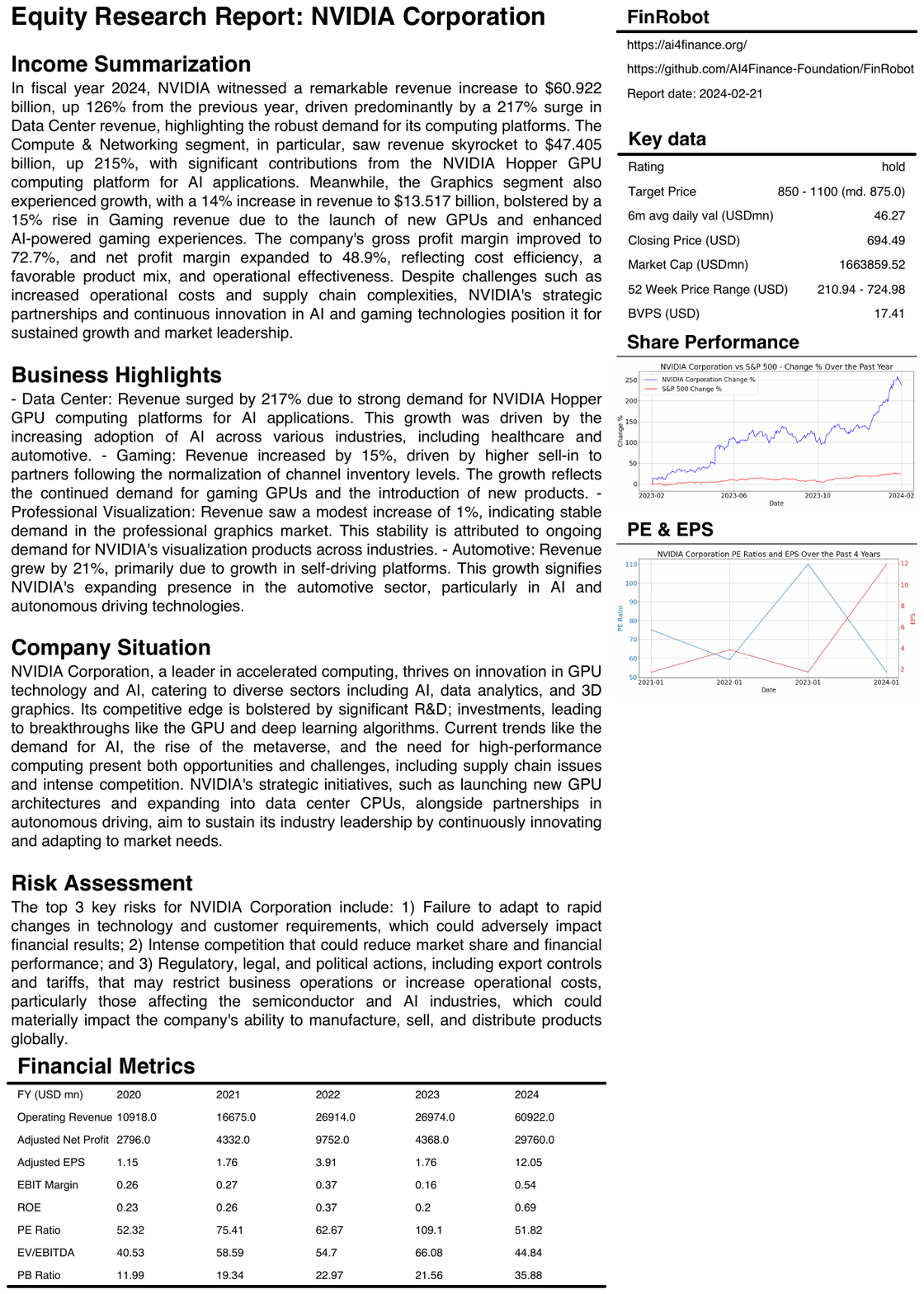}
\label{fig:nvda_report1}
\vspace{-1mm}
\end{figure*}

\begin{figure*}
\centering
\includegraphics[scale = 0.77]{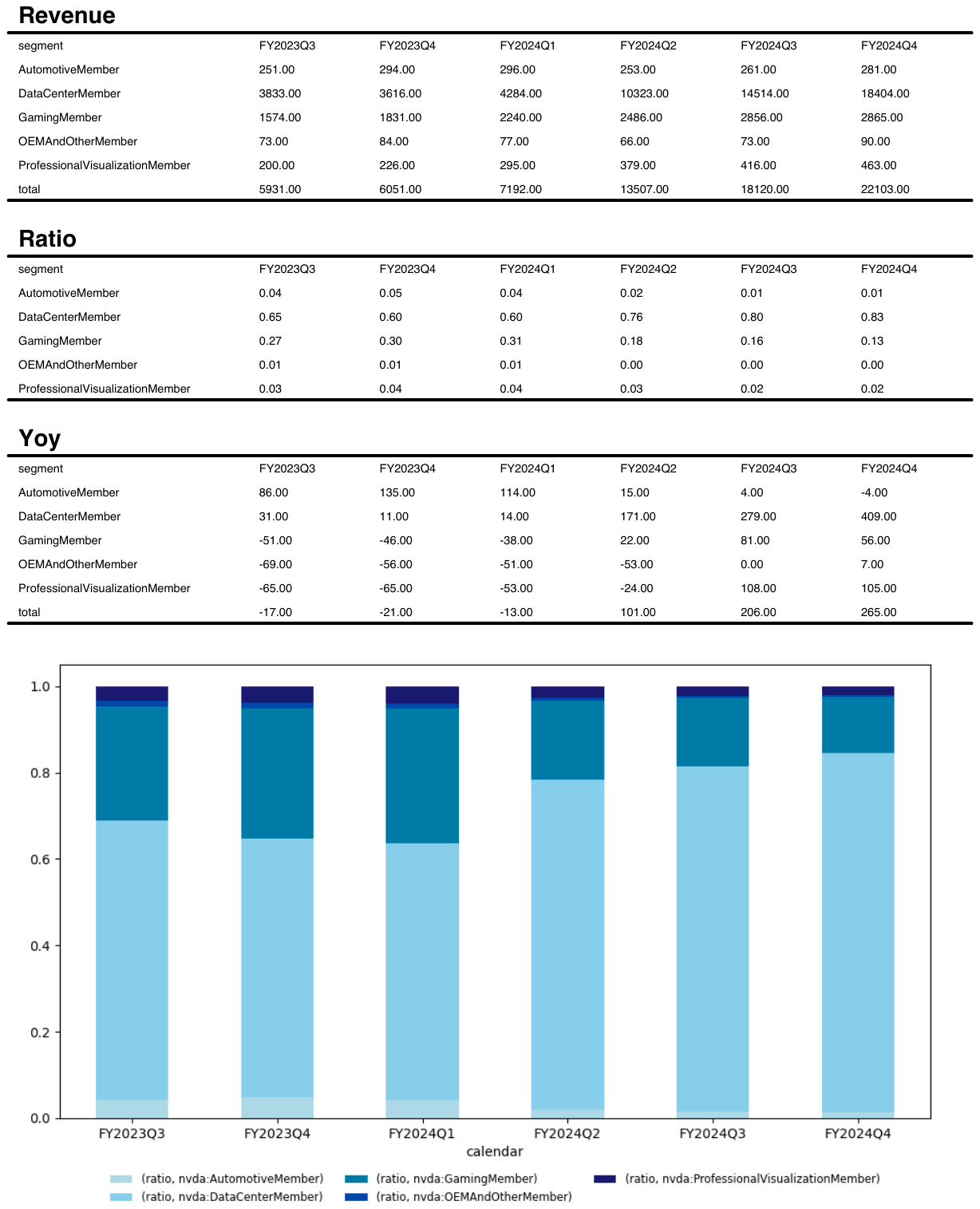}
\label{fig:nvda_report2}
\vspace{-1mm}
\end{figure*}

\begin{figure*}
\centering
\includegraphics[scale = 0.77]{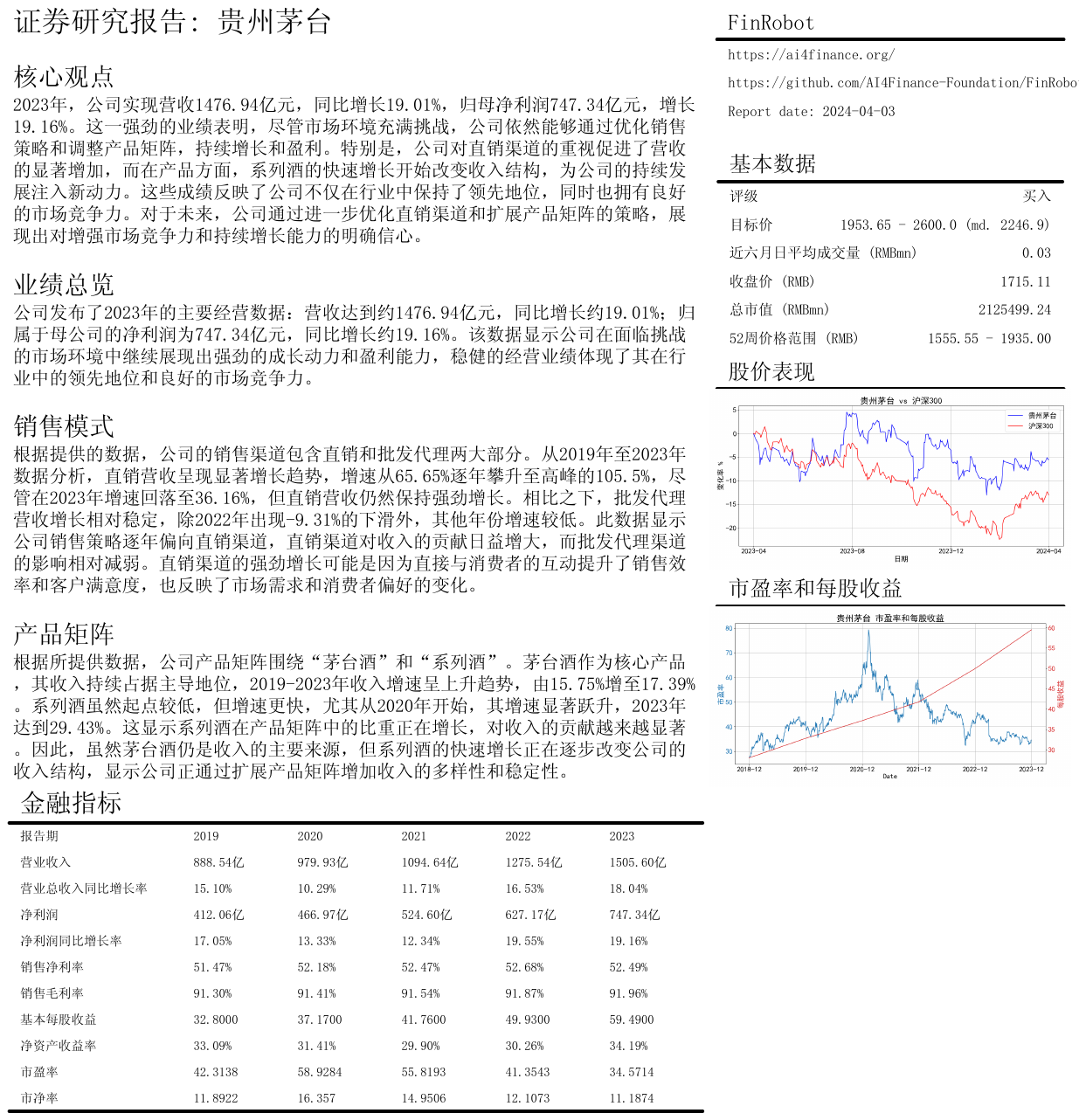}
\label{fig:maotai_report1}
\vspace{-1mm}
\end{figure*}

\begin{figure*}
\centering
\includegraphics[scale = 0.77]{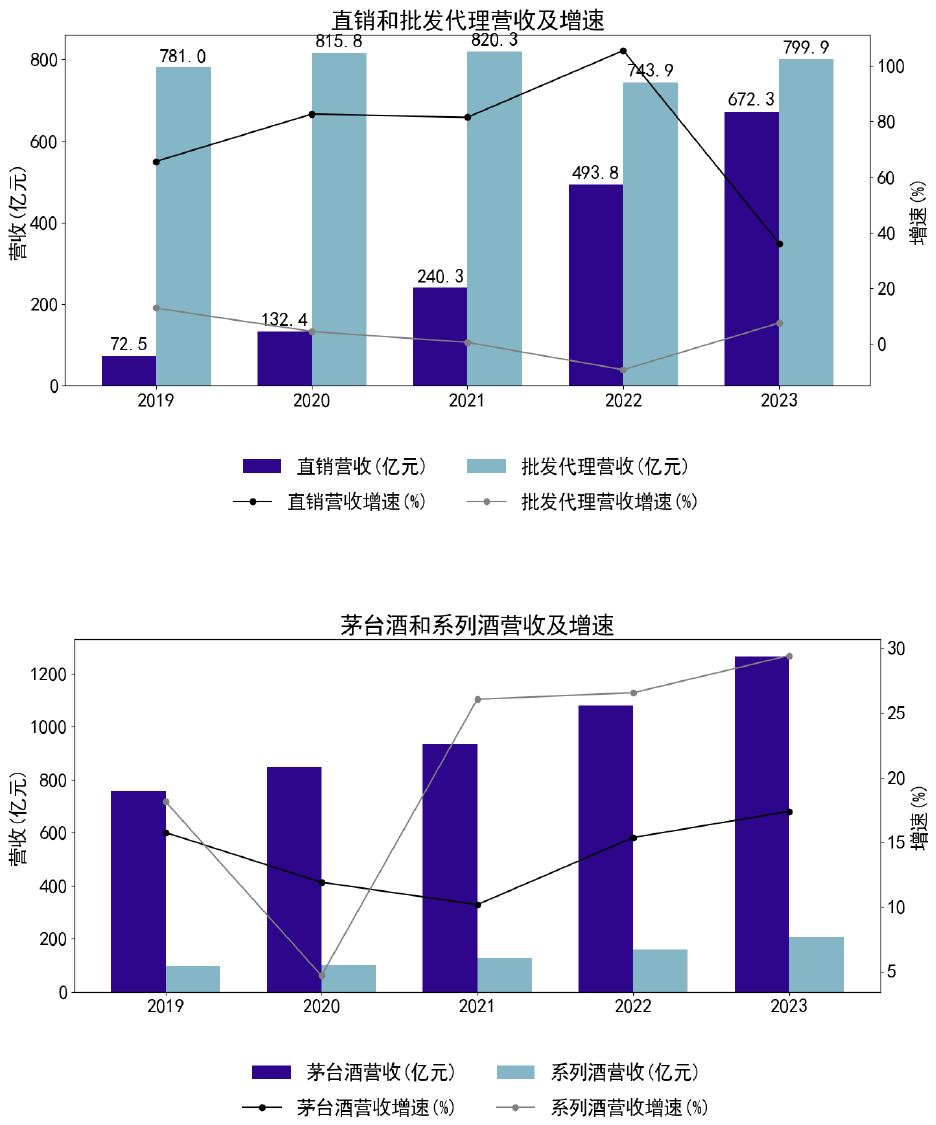}
\label{fig:maotai_report2}
\vspace{-1mm}
\end{figure*}

\end{document}